\documentstyle[12pt]{article}
\def\beq{\begin{equation}}
\def\eeq{\end{equation}}
\def\bea{\begin{eqnarray}}
\def\eea{\end{eqnarray}}
\def\bq{\begin{quote}}
\def\eq{\end{quote}}

\def\nnb{\nonumber}
\def\ga{\left(}
\def\dr{\right)}
\def\aga{\left\{}
\def\adr{\right\}}

\def\rar{\rightarrow}
\def\nnb{\nonumber}
\def\la{\langle}
\def\ra{\rangle}
\def\nin{\noindent}
\def\ba{\begin{array}}
\def\ea{\end{array}}

\begin{document}
\topmargin -1.5cm
\oddsidemargin -.5cm
\evensidemargin -1.0cm
\pagestyle{empty}
\begin{flushright}
PM 95/41\\
hep-ph/9510212\\
(to appear as Phys. Lett. B)
\end{flushright}
\vspace*{0.5cm}
\begin{center}
\section*{$B^*B\pi(\gamma)$ couplings and $D^*\rar D\pi(\gamma)
$-decays
\\ within a $1/M$-expansion in $full$ QCD}
%the $M_b \rar \infty$ limit
%\\from QCD
%spectral moment sum rules}
\vspace*{0.5cm}
{\bf H.G. Dosch \footnote{On leave of absence
from Institut Theoretical Physics of the University of
Heidelberg.} and S. Narison} \\
\vspace{0.3cm}
Laboratoire de Physique Math\'ematique\\
Universit\'e de Montpellier II\\
Place Eug\`ene Bataillon\\
34095 - Montpellier Cedex 05, France\\
\vspace*{1.2cm}
{\bf Abstract} \\ \end{center}
%\vspace*{2mm}
\nin
To leading order in $\alpha_s$, we evaluate the leading and
non-leading $1/M_b$ corrections
to the $B^*B\pi$ and
 $B^*B\gamma$
couplings using QCD
spectral moment sum rules in the full theory.
We find that,
for large $M_b$ and
contrary to the heavy-to-light $B\rar \pi(\rho) l\bar\nu$ form
factors,
which are dominated by the $soft$
light quark vacuum condensate,
these couplings are governed by the $hard$ perturbative
graph, like other heavy-to-heavy
transitions.
We also find that
for the $B^{*}\rar B\gamma$, the $1/M_b$
correction is mainly due to the
perturbative and light quark condensate contributions
originating from the graphs involving the heavy
quark part of the electromagnetic current, which
are essential for explaining the large charge dependence in
the observed $D^{*-}\rar D^-\gamma$ and
$D^{*0}\rar D^0\gamma$ decays. Our $best$ numerical predictions
{\it without any free parameters} for the
$B^*$-meson are:
$g_{B^{*-}B^0\pi^-}\simeq 14\pm 4$, $
\Gamma_{B^{*-}\rar B^-\gamma}\simeq (0.10\pm 0.03)$ keV
and the large charge dependence of the ratio:
${\Gamma_{B^{*-}\rar B^- \gamma}}/
{\Gamma_{B^{*0}\rar B^0 \gamma}}\simeq 2.5~$
. For the $D^*$-meson, we find:
$\Gamma_{D^{*-}\rar D^0\pi^-}\simeq 1.54\Gamma_{D^{*0}\rar
D^0\pi^0}
\simeq (8\pm 5)$ keV,
$\Gamma_{D^{*-}\rar D^-\gamma}\simeq (0.09^{+0.40}_{-0.07} )$
keV and
$\Gamma_{D^{*0}\rar D^0\gamma}\simeq (3.7\pm 1.2)$ keV,
where the branching ratios agree within the errors with the
present data, while
the total widths $\Gamma_{D^{*0}\rar all}\simeq (11\pm 4)$ keV
and $\Gamma_{D^{*-}\rar all}\simeq (12\pm 7)$ keV are much
smaller
than the present experimental upper limits.
\vspace*{2.cm}
\begin{flushleft}
PM 95/41\\
hep-ph/9510212\\
%\today
September 1995
\end{flushleft}
\vfill\eject
\setcounter{page}{1}
 \pagestyle{plain}
\section{Introduction and notations} \par
The  $B^*B\pi$ and $D^*D\pi$ couplings have been studied by
several authors
using
QCD spectral sum rules combined with the soft pion
techniques \cite{GATTO}(see also \cite{ELET}),
light cone sum rules \cite{BRAUNC} or
heavy quark expansion plus soft pion techniques \cite{BURD},
while the
$B^*B\rho$ coupling has been studied recently using QCD double
exponential sum rules for the three-point function \cite{PAK}.
\nin
However, though, apparently convenient, as one works with the
two-point function, the sum rule approach of \cite{GATTO} is
quite peculiar
due to the presence of the unphysical so-called $parasitic~
term$,
which can only be eliminated in a more involved combination
of sum rules. Moreover, the connection of the light-cone sum
rule
used in \cite{BRAUNC},
with the $standard$ QCD spectral sum rules (QSSR) involving the
vacuum condensates, is not crystal clear due to the poor
understanding in terms of the vacuum condensates
of the real structure of the light meson wave
functions entering into this approach (for a criticism on the
unreliability for the construction of the hadronic wave
functions on the light cone, see e.g. \cite{ECK}).
\nin
The theoretical evaluations of these couplings are
interesting as they can be used for
determining the normalization of the $B\rar \pi (\rho)~l\nu$
form
factors near zero pionic recoil.
Indeed, the QSSR analysis of the $q^2$-dependence
of the $B\rar \pi l\bar\nu$ form factor
$f_+$ indicates that it behaves like
a polynomial in $q^2$, which can be fitted to a
good approximation with the pole form \cite{DOSCH,SN1},
though one cannot use (like currently done in the literature)
such a pole
parametrization for studying its $M_b$-behaviour at $q^2=0$
\cite{SN1}.
\nin
Experimental measurements of these couplings
are expected to be improved and
available in the forthcoming high-statistics
$B$- and $\tau$-$charm$-factory machines
from the processes $B\rar\pi\tau\nu$,
$D^*\rar D\pi$ and $B^*(D^*)\rar B(D)\gamma$.
\nin
In this paper, we shall use the
 QSSR  double moments sum rule approach in order to
study the large
$M_b$-behaviour of the previous couplings and to estimate their
values. Contrary to the
$popular$ double exponential sum rule,
this approach is quite advantageous in the
analysis of the three-point function,
as it prevents the blow-up of the
QCD series when the heavy quark mass is large but here the
number of derivatives remains finite  \cite{SN1}.
\nin
The couplings are defined as:
\beq
\la B^*(p)B(p')\pi(q)\ra = g_{B^*B\pi}q_\mu \epsilon^\mu ,
 ~~~~~~~~~
\la B^*(p)B(p')\gamma(q)\ra = -eg_{B^*B\gamma}p_\alpha
p'_\beta \epsilon^{\mu\nu\alpha\beta}\epsilon_\mu
\epsilon'_\nu~,
\eeq
where $q \equiv p'-p$ and $-Q^2\equiv q^2\leq 0 $, while
$\epsilon_\mu$ are the polarization of the vector particles. We
shall be concerned with the vertex function:
\beq
V^{\nu (\mu)} (p,p',q)=-\int d^4x~ d^4y~e^{i(p'y-px)}\la
{\cal T}
J_L^{5(\mu)}(x)J^\nu_{B^*} (0) J_B(y) \ra~,
\eeq
where the currents are:
\bea
J^{\mu}_L&=& \sum_{u,d}e_q\bar q \gamma^\mu q+
\sum_{c,b}e_Q\bar Q \gamma^\mu Q~~~~~~~~~
J^{5}_L= (m_u+m_d)\bar u \gamma^5 d \nnb\\
J^\nu_{B^*}&=&\bar u \gamma^\nu b ~~~~~~~~~
J_B= (M_b+m_d)\bar d \gamma^5 b~,
\eea
 and $ u,~d,~c,~b$ are the quark
fields and $e_q,~ e_Q$ their electric charge in units of $e$.
The vertex obeys the double dispersion relation \footnote
{Here, the dispersion relation is done with
respect to the two heavy meson momenta, which is not the case of
the $B\rar \pi (\rho) l\bar\nu$. This different configuration
is important for the $M_b$-behaviour of the
different QCD contributions.}:
\beq
V(p,p',q)= -\frac{1}{4\pi^2}\int_{M_b^2}^{\infty}\frac{ds}
{s-p^2}
 \int_{M_b^2}^{\infty}\frac{ds'}{s'-p'^2}~
\mbox{Im}V(s,s')~+...
\eeq
Exploiting the fact that $M_b$ is much larger than the QCD scale
$\Lambda$, where the LHS can be evaluated
using the Operator Product Expansion \`a la SVZ \cite{SVZ,SNB},
one can work, in the chiral limit, with the double moment sum
rule:
\beq
{\cal M}^{(n,n')}= -\frac{1}{4\pi^2}\int_{M_b^2}^{\infty}
\frac{ds}{s^{n+1}}
 \int_{M_b^2}^{\infty}\frac{ds'}{s'^{n'+1}}~
\mbox{Im}V(s,s')
\eeq
where $n,~n'$ are $finite$
numbers of derivatives evaluated at $p^2=p'^2
=0$.
\section{Moment sum rule for the $B^*B\pi$ and $D^*D\pi$
couplings}
The perturbative QCD expression of the spectral function reads:
\beq
-\frac{1}{4\pi^2}\mbox{Im}V(s,s')=(m_u+m_d)M_b\frac{N_c}
{4\pi^2}
Q^2
\frac{M^2_b(s+s'+Q^2)-2ss'}{\aga (s+s'+Q^2)^2-4ss'\adr^{3/2}}~,
\eeq
where $Q^2\equiv -q^2 \geq 0$
\footnote{In this region of $Q^2\geq 0$, the
question of
non-Landau and complex singularities and of anomalous
thresholds do not arise \cite{DOSCH2}.}
is the pion momentum squared and
where the integration limit condition is:
\beq
(s-M^2_b)(s'-M^2_b)\geq Q^2M^2_b.
\eeq
It is easy to
check that, contrary to the case of $B \rar \pi l\bar\nu$
form factor, where
the light quark condensate is dominant,
the light condensate contribution vanishes here after
taking the $p^2$ and $p'^2$ derivatives, which is a
consequence of the
fact that the dispersion relation has been done with respect
to the
heavy quarks momenta like in the case of a heavy-to-heavy
transition.
The other remaining
effects which are suppressed by 1/$M^2_b$ compared to the
leading
perturbative diagram will be neglected to the approximation
we are working \footnote{One can however notice that
the four-quark condensate contribution behaves like $1/Q^4$
which reflects the fact that the present approach cannot be
used at
$Q^2=0$ as expected.}.
\nin
The phenomenological side of the sum rule
is parametrized using the usual duality ansatz:
{\it lowest resonance}
+{\it QCD continuum} from the thresholds $s_c$ and $s'_c$. By
transferring this QCD continuum effect into the QCD part of the
sum rule, one obtains \footnote{Here and in the
following, we shall neglect the contribution of the $\pi'$(1.3) similarly
to previous analysis of the $\omega\rho\pi$- and
$\pi NN$-couplings using
vertex sum rules \cite{SNB}.}:
\beq
{\cal M}^{(n,n')}_c\equiv
g_{B^*B\pi}\frac{\sqrt{2}M_{B^*}f_{B^*}}{M^{2(n+1)}_{B^*}}
\frac{\sqrt{2}M^2_{B}f_{B}}{M^{2(n'+1)}_{B}}
\frac{\sqrt{2}m^2_\pi f_\pi}{m^2_\pi+Q^2}\simeq
-\frac{1}{4\pi^2}\int_0^{s_c}\frac{ds}{s^{n+1}} \int_0^{s'_c}
\frac{ds'}{s'^{n'+1}}
{}~\mbox{Im}V(s,s'),
\eeq
where the coupling constants are normalized as:
\bea
\la 0|J^5_L|\pi\ra&=&\sqrt{2}f_\pi M^2_\pi~~~~~
\la 0|J^\mu_L|\rho\ra=\sqrt{2}\frac{M^2_\rho}{2\gamma_\rho}
\epsilon^\mu,
\nnb\\
\la 0|J_B|B\ra&=&\sqrt{2}f_BM^2_B~~~~~
\la 0|J_{B^*}^\mu |B^*\ra=\sqrt{2}f_{B^*}M_{B^*}\epsilon^\mu,
\eea
where $f_\pi=93.3$ MeV and $\gamma_\rho= 2.56$.
In the case where $M_b\rar \infty$ (static limit),
it is convenient to
 work with the
non-relativistic variables $E$ and $E'$ defined as:
\beq
s= (E+M_b)^2~~~~~~~\mbox{and}~~~~~~~~s'= (E'+M_b)^2,
\eeq
and to introduce the new variables:
\beq
x=E-E'~~~~~~~~~~~~~~~\mbox{and}~~~~~~~~~~~~~~~
y=\frac{1}{2}(E+E')~.
\eeq
Due to the almost good symmetry between the $B$ and the $B^*$,
we shall use:
\beq
M_{B^*}\simeq M_B,~~~~~~~~ E_c\simeq E'_c,
{}~~~~~~~~n=n'\equiv n_3~.
\eeq
By keeping the non-leading 1$/M_b$-terms in the expansion, we
obtain to leading order in $\alpha_s$:
\bea
{\cal M}^{(n,n')}_c&\simeq &(m_u+m_d)
\frac{M^3_b}
{M_b^{4(n_3+1)}}\frac{N_c}{\pi^2}
Q^2  \int_0^{E_c}dx\int_{
\frac{1}{2}\sqrt{x^2+Q^2}}^{E_c-\frac{x}{2}} \frac{y~dy}
{\ga x^2+Q^2
\dr^{3/2}}
\Bigg{[} 1+\nnb\\
&&\frac{1}{M_b}\Big{[}-\frac{1}{4y}(Q^2+\frac{3}{2}x^2
+2y^2)-2(2n_3+1)y +
\frac{Q^2}{8M_b}\delta(y-
\frac{1}{2}\sqrt{x^2+Q^2})\Big{]}\Bigg{]}~.
\eea
For consistency, we shall use in our analysis the lowest
order expression in $\alpha_s$
of the decay constants from the moment
sum rules \cite{SNE}:
\bea
f^2_B &\simeq&
 \frac{E^3_c}{2\pi^2}\frac{1}{M_B}
\ga \frac{M_B}{M_b}\dr^{2n_2-1}
\aga 1-\frac{3}{2}(n_2+1)\frac{E_c}{M_b}-\frac{\pi^2}{2}
\frac{\la \bar dd \ra }{E^3_c}\adr~\nnb\\
f^2_{B^*}
&\simeq&\frac{E^3_c}{2\pi^2}\frac{1}{M_B}
\ga \frac{M_B}{M_b}\dr^{2n_2+3}
\aga 1-\frac{3}{2}(n_2+\frac{7}{3})\frac{E_c}{M_b}-\frac{\pi^2}
{2}
\frac{\la \bar dd \ra }{E^3_c}\adr~,
\eea
consistent with the normalization of the currents in Eq.(3) and with
the definitions in Eq.(9).
One should notice that the overall $(M_B/M_b)$ factor also
brings
a $1/M_b$-correction which tends to reduce the $apparently$
huge
correction in the curly
brackets and leads after the moment sum rules analysis of $f_B$ to the
well-known 1 GeV/$M_b$-correction to this quantity.
One can also notice that the $1/M_b$-correction to $f_{B^*}$ is slightly
smaller than the one of $f_B$ as generally expected.
Using the previous formulae in
Eq. (14), the emerging
{\it effective} values of $E_c$ fixed from the numerical
analysis
of $f_B$ and $f_{B^*}$
including the $\alpha_s$ corrections are \cite{SNE,ZAL}:
\beq
E^{\infty}_c\simeq (1.6\pm 0.1)~\mbox{GeV}~~~~~
E^{B}_c\simeq (1.3\pm 0.1)~\mbox{GeV}~~~~~
E^{D}_c\simeq (1.1\pm 0.2)~\mbox{GeV}~,
\eeq
which, using Eq. (10), can be parametrized as:
\beq
E_c=E_c^{\infty}\ga 1-\frac{E_c}{2M_b}\dr .
\eeq
As in \cite{SN1}, we minimize the $n$-dependence of the results by
requiring that the leading term is $n$-independent. This leads
to
the constraint:
\beq
4n_3+1=2n_2+1~,
\eeq
where $n_2\simeq 4-5$ is the value where $f_B$ from
the two-point function has been optimized \cite{SNE}.
By
evaluating $numerically$ the different integrals, we obtain to a
good approximation:
\beq
g_{B^*B\pi}\simeq g^{LO}_{B^*B\pi}\aga
1+\frac{3}{2}\frac{E_c}{M_b}+
\frac{\pi^2}{2}
\frac{\la \bar uu \ra }{E^3_c}\adr
\eeq
where:
\beq
 g^{LO}_{B^*B\pi}\simeq
\frac{N_c}{\sqrt{2}f_\pi} \frac{m_u+m_d}{m^2_\pi}
M_B \aga {\cal I}_0\equiv \frac{Q^4}{E_c^3}
\int_0^{E_c}dx\int_{
\frac{1}{2}\sqrt{x^2+Q^2}}^{E_c-\frac{x}{2}} \frac{y~dy}
{\ga x^2+Q^2
\dr^{3/2}}\adr~.
\eeq
\nin
The analytic expression of the integral ${\cal I}_0$ is:
\beq
{\cal I}_0(\rho\equiv Q/E_c)=\frac{Q}{2}\rho
\Bigg{[} \frac{1+\frac{3}{4}\rho^2}{\sqrt{1+\rho^2}}-\rho
\Bigg{]}~,
\eeq
which exhibits a broad maximum
in the range 1--3 GeV$^2$. At $Q^2\simeq 2$ GeV$^2$,
where the absolute maximum is obtained, its numerical value for different
$E_c$ can be parametrized by the interpolating formula:
\beq
{\cal I}_0\simeq (0.119\pm 0.001)E_c~.
\eeq
Therefore, using Eq. (16), we finally obtain:
\footnote{We have checked using the complete perturbative
expressions of the three- and two-point functions that the
higher order
$1/M$ corrections are small and do not spoil the validity of
the following
approximate formula even at the charm mass.}
\beq
g_{B^*B\pi}\simeq \frac{2M_B}{\sqrt{2}f_\pi}g^{\infty}
\aga
1+\frac{E^B_c}{M_b}+
\frac{\pi^2}{2}
\frac{\la \bar uu \ra }{(E^B_c)^3}\adr~,
\eeq
where we have introduced the static coupling $g^{\infty}$:
\beq
g^{\infty}\equiv \frac{N_c}{2}
\ga\frac{m_u+m_d}{m^2_\pi}\dr
\ga 0.119E^{\infty}_c\dr~,
\eeq
 which
controls the interaction of the pion with infinitely heavy
fields in the effective Lagrangian approach:
\beq
{\cal L}_{\mbox{int}}=\frac{i}{2}g^{\infty}{\bf Tr}
H\gamma^\mu\gamma_5(\pi^{\dagger}\partial_\mu\pi
-\pi\partial_\mu\pi^{\dagger})\bar H,
\eeq
where $H$ and $\pi$ are the heavy and pion fields.
The $M_b$-behaviour obtained here, which is
dictated by the one of $f^2_B$ is in agreement with
current expectations \cite{GATTO}-\cite{BURD}. The agreement
with
the one in \cite{GATTO} (the sum rule used in \cite{GATTO} is
very similar to the light-cone sum rule in the treatment of
the pion)
 and the light-cone sum rule \cite{BRAUNC}
can be mainly due to the fact that, in the present process,
the $hard$ perturbative diagram gives the $leading
{}~contribution$
in $1/M_b$,
where the present version of the light-cone sum rule approach,
which is dominated,
(by construction),
by the hard perturbative diagram, is
appropriate. In the case where  the $soft$ process is
dominant, like e.g. in the analysis of the $B\rar \pi(\rho)$
semi-leptonic processes \cite{SN1},
one should need a modified version of the light-cone sum rule
approach in order to take properly into account the dominant
non-perturbative $\la\bar uu \ra $ condensate contribution.
\nin
We expect that, in the
moment sum rules , the $\alpha_s$
correction is much smaller than
the one in the non-relativistic
exponential one, as here $\alpha_s$
is evaluated at a larger scale of about $M_b/\sqrt{n}$, while
in the exponential sum rule the scale is much lower at about 1
GeV.
Therefore,
we expect that the expression in Eq. (22) with the value
of $E_c$ in Eq. (15) gives a good
approximation of the $physical$ result.
However, we consider, as an intrinsic error of the approach,
the known
30\% effect due to $\alpha_s$ in the decay constants
from the moment sum rules \cite{SNE}.
We shall use $(\overline{m}_u+\overline{m}_d)$ (1GeV)
=(12.5$\pm$2.5) MeV \cite{MASS} rescaled at $Q^2=2$ GeV$^2$,
and the corresponding value of the quark condensate.
Then, we deduce:
\beq
g^{\infty}\simeq (0.15 \pm 0.03).
\eeq
where the error takes into account the effect of 30\%
by the radiative correction to the value of $f_B$ \cite{SNE}.
\nin
Our prediction in Eq. (25) is in agreement with
the range of values obtained in \cite{GATTO}, though the authors
in \cite{GATTO} use
a too low value of $f_B^{static}\simeq 1.45f_\pi$, while
we use here the two-loop
value $f_B^{static} \simeq 2f_\pi$ from \cite{SNE}
and from the recent lattice
results \cite{ALLTON}.
\nin
For the $physical~B$-meson, and including the $1/M_b$
correction which is of the order of +28 \% at the $b$-mass,
we obtain:
\beq
g_{B^*B\pi}\simeq 14.5 \pm 3.3~,
\eeq
where again the error takes into account the effect of
radiative corrections to $f_B$. We have used the
two-loop {\it non-relativistic pole} masses \cite{SNM}:
\beq
M_b= (4.7\pm 0.03)~\mbox{GeV}~~~~~~~~
M_c= (1.45\pm 0.05)~\mbox{GeV},
\eeq
consistent with the present use of non-relativistic sum rules.
 One can notice
that, like in the case of $f_B$, the
$1/M_b$-correction is
large (28\% at the $b$-quark mass and 76\% at the
$c$-quark mass). This feature can make the extrapolation of
the result to the $D$-meson quite
risky. However, as already mentioned earlier, our explicit
evaluation of the complete perturbative expression of the
three- and two-point correlators indicate that higher order
corrections
in 1/$M$ remain small. Therefore, we can deduce, with
a quite good confidence, the estimate from the
moments:
\beq
g_{D^*D\pi}\simeq 7.1\pm 1.6~.
\eeq
We cross check the validity of the previous result by invoking
semi-local duality sum rules for
the two- or three-point functions
which correspond respectively to the particular cases where
$(n_3,n_2)=(-1/2,-1)$ or $(n_3,n_2)=(-1,-2)$,
and which have been discussed extensively in the
case of the QCD two-point functions for light \cite{BERTL} and
heavy quarks \cite{PICH,ZAL}.
In these particular cases, the $1/M$ corrections to the
leading term of
the three-point function are much
smaller and result by a correction of about +(9-11)\% and +
(23-26)\%
for the coupling respectively at the $b$ and $c$ quark masses,
giving
\footnote{One should also notice that the use of the Laplace
sum rules leads to a small value of the optimization scale
$\tau$, which is $pratically$ similar to the semi-local duality
sum rules used here.}:
\beq
g_{B^*B\pi}\simeq 12.7 \pm 2.9~,~~~~~~~g_{D^*D\pi}\simeq
5.0\pm 1.1~.
\eeq
By considering the previous
results, we conclude that the {\it most conservative estimate}
 of the couplings from the sum rule is:
\beq
g_{B^*B\pi}\simeq 14 \pm 4~,~~~~~~~g_{D^*D\pi}\simeq 6.3\pm
1.9~.
\eeq
Our final results are in good agreement with the ones in
\cite{GATTO} but are much smaller than the ones from
\cite{BRAUNC} and to the indirect determination of \cite{LELOU},
which is correlated to a higher input value of the $B\rar \pi
l\nu$ form factor.
Taking into account that the $1/M_b$-correction to $f_{B{(^*)}}$
is approximately equal in strength but opposite in sign
with the one for $g_{B^*B\pi}$, one obtains with a good
accuracy:
\beq
R_{BD}\equiv \frac{g_{B^*B\pi}f_{B^*}\sqrt{M_B}}
{g_{D^*D\pi}f_{D^*}\sqrt{M_D}}\simeq 1~,
\eeq
as expected from an alternative analysis \cite{BURD}.
 Our numerical values of the $physical$
couplings are in better agreement with the results
in \cite{GATTO} including the radiative
corrections  than with the ones in \cite{BRAUNC}, which
is higher than ours by a factor 2 . \footnote
{The agreement with \cite{GATTO}
for the $physical~B$-meson is due to the approximately same
value
of $f_B$ used here ($1.5f_\pi$) and in
\cite{GATTO} ($1.36f_\pi$), though we, orginally, do not start
with the
same value of $f_B^{static}$. In our analysis,
the value of $f_B^{phys}\simeq 1.5f_\pi$,
at the $physical~B$-meson mass has been deduced from $f_B^{stat}
\simeq 2f_\pi$,
after taking into account the $1/M_b$-correction.
In Ref. \cite{GATTO}, this $1/M_b$-effect seems to be
much smaller than currently expected as, there, $f_B^{stat}
\simeq
1.45 f_\pi\approx f_B^{phys}$.}
The coupling in Eq. (30) leads to the prediction:
\beq
\Gamma_{D^{*-}\rar D^0\pi^-}=\frac{g_{D^*D\pi}^2}{24\pi
M^2_{D^*}}
|\vec{q}_\pi|^3\simeq
1.54\Gamma_{D^{*0}\rar D^0\pi^0}\simeq (8\pm 5)~\mbox{keV}~,
\eeq
where we have assumed isospin invariance for the couplings.
 Using the observed branching ratios \cite{PDG},
 one can also predict the radiative decays:
\beq
\Gamma_{D^{*0}\rar D^0 \gamma}\simeq (3.0\pm 1.2)~\mbox{keV}
{}~~~~~~~~~~~~~~~
{\Gamma_{D^{*-}\rar D^- \gamma}}\simeq .13^{+.33}_{-.11}~\mbox
{keV}~.
\eeq
and the total widths:
\beq
\Gamma_{D^{*-}\rar all}\simeq (12\pm 7)  ~\mbox{keV}
{}~~~~~~~~~~~~~~
\Gamma_{D^{*0}\rar all}\simeq (8\pm 3) ~\mbox{keV}~.
\eeq
The predictions for the total widths
are much smaller than the present experimental
upper limits.
An improved measurement of the $D^*$-total widths in
the next tau-charm factory machine should provide a decisive
test for the validity of these extrapolated predictions.
\section{Moment sum rule for the $B^*B\gamma$ and
$D^*D\gamma$ couplings}
The QCD expression can be decomposed
into a light $(q\equiv
u,~d,~s)$ and
heavy $(Q\equiv c,~b)$ quark parts.
For the corresponding vertex function, the one
available in \cite{PAK} agrees with our recomputation apart for
the relative sign between the perturbative and quark condensate
contributions in the heavy quark component of the
electromagnetic
cuurent.  After a systematic
$1/M_b$-expansion
of the full QCD expression, one can inspect that the dominant
contribution comes from the perturbative graph related to the
light quarks coupled to the electromagnetic current. The
heavy quark contribution is $1/M_b$-suppressed compared to the
light quark one. However, the perturbative and
light quark condensate contributions are of the same order in
$1/M_b$ in this heavy quark component. By
keeping the 1$/M_b$-correction, the QCD part of the sum rule
reads:
\bea
{\cal M}^{(n,n')}_c \Big{\vert}_{QCD}&\simeq &
e_q \frac{M_b}{M_b^{4(n_3+1)}}\Bigg{[}
M_b\frac{N_c}{\pi^2}
Q^2  \int_0^{E_c}dx\int_{
\frac{1}{2}\sqrt{x^2+Q^2}}^{E_c-\frac{x}{2}} \frac{y ~dy}
{\ga x^2+Q^2
\dr^{3/2}}
\Bigg{[} 1+\nnb\\
&&\frac{1}{M_b}\Big{[}\frac{1}{4y}(Q^2+\frac{x^2}{2}
-10y^2)-2(2n_3+1)y +
\frac{Q^2}{8M_b}\delta(y-
\frac{1}{2}\sqrt{x^2+Q^2})\Big{]}\Bigg{]}\nnb\\
&-&
\frac{16}{9}\pi
\frac{\alpha_s\la \bar uu \ra^2}
{Q^4}\Bigg{]}+\frac{e_Q}{M_b^{4(n_3+1)}}\Bigg{[}
\frac{N_c}{3\pi^2}E^3_c-{\la\bar uu\ra}+{\cal{O}}(1/M_b)
\Bigg{]}~,
\eea
where $e_{q(Q)}$ is the charge of the light (heavy) quark in
units of $e$.
The phenomenological side of the sum rule can be parametrized
as:
\beq
{\cal M}^{(n,n')}_c\Big{\vert}_{phen}\simeq
g_{B^*B\gamma}\frac{\sqrt{2}M_{B^*}f_{B^*}}{M^{2(n_3+1)}_{B^*}}
\frac{\sqrt{2}M^2_{B}f_{B}}{M^{2(n_3+1)}_{B}}~.
\eeq
Using an approach similar to the one done for $B^*B\pi$,
we deduce the sum rule:
\beq
g_{B^*B\gamma}(Q^2)\equiv g^L_{B^*B\gamma}(Q^2)+g^H_{B^*B
\gamma}(Q^2)
\eeq
where:
\bea
g^L_{B^*B\gamma}(Q^2)&\simeq& e_q
\Bigg{[}\frac{N_c{\cal I}_0}{Q^2}
\ga 1+\frac{1}{2}\frac{E^B_c}{M_b}\dr
-\frac{16}{9}\pi
\frac{\alpha_s\la \bar uu \ra^2}
{(E^B_c)^3Q^4} \Bigg{]}~,\nnb\\
g^H_{B^*B\gamma}(Q^2)&\simeq&
\frac{e_Q}{M_b}
\Bigg{[} \frac{N_c}{3}-\frac{\pi^2\la\bar uu\ra}
{\ga E^B_c\dr^3}\Bigg{]}~,
\eea
where ${\cal I}_0$ is the integral defined
in Eq. (19).
For $M_b\rar\infty$, the coupling is given by the light
quarks contribution and remains constant. The $1/
M_b$-correction
in the light quark contribution is much smaller
than the one for $g_{B^*B\pi}$ since there
is an almost cancellation of the $1/M_b$-correction with the
one from the $E_c$-dependence of ${\cal I}_0$ as can be
deduced from Eq. (16), while the one due to the heavy
quark is important at the $c$-quark mass. The light quark
coupling exhibits a typical monopole behaviour
 for $ Q^2\geq M^2_\rho~$, while
the heavy quark coupling is $Q^2$-independent. Therefore, we use
a light vector meson dominance for the estimate of the light
quark coupling, which can be related to the $B^*BV$-coupling
($V\equiv \rho,~\omega$) as:
\beq
g^L_{B^*B\gamma}(Q^2)=\ga\frac{\sqrt{2}M^2_V}{2\gamma_V}\dr
\frac{e_q}{Q^2+M^2_V}g^L_{B^*BV}~.
\eeq
A sum rule analysis of the $B^*BV$-coupling
similar to the one for $B^*B\pi$, shows a very good stability
for
$0.4\leq Q^2\leq 2.2$ GeV$^2$, where the optimal value obtained
for $Q^2\approx 1$ GeV$^2$ reads:
\beq
g^L_{B^*BV}\simeq (0.84\pm 0.10)
/\ga\frac{\sqrt{2}M^2_V}{2\gamma_V}\dr
\eeq
We have used ${\alpha_s\la \bar uu \ra^2}\simeq (5.8\pm 0.9)10^
{-4}
$ GeV$^4$ \cite{SNP,SNB}, which shows a negligible
contribution of
the four-quark condensate in the range of $Q^2$-stability.
Therefore, we deduce:
\bea
g_{B^*B\gamma}(Q^2=0)
 &\simeq&\Bigg{[} e_q (1.14\pm 0.15)+e_Q\frac{(0.90\pm 0.16)
{}~\mbox{GeV}}{M_b}\Bigg{]} ~\mbox{GeV}^{-1}~,\nnb\\
g_{D^*D\gamma}(Q^2=0)
 &\simeq&\Bigg{[} e_q (1.11\pm 0.24)+e_Q
\frac{(0.90\pm
0.16)~\mbox{GeV}}{M_c}\Bigg{]}~\mbox{GeV}^{-1}~.
\eea
For the $B$-meson, the heavy quark contribution is relatively
small. One obtains:
\beq
\Gamma_{B^{*-}\rar B^- \gamma}= g_{B^*B\gamma}^2
\frac{\alpha}{3}|\vec{q}_\gamma|^3\simeq (.10\pm .03)~
\mbox{keV}~.
\eeq
and with a better accuracy for the ratio:
\beq
\frac{\Gamma_{B^{*-}\rar B^- \gamma}}
{\Gamma_{B^{*0}\rar B^0 \gamma}}\simeq 2.5~,
\eeq
which deviates strongly from the na\"{\i}ve static
limit ($M_b \rar \infty$) expectation $(e_u/e_d)^2=4$.
For the $D$-meson, the heavy quark contribution is relatively
important. One should notice that
the $\Upsilon$-contribution has been completely ignored in the
phenomenological analysis of \cite{PAK}, which
can explain their opposite prediction of this ratio
with respect to us and to the data.
We deduce within the previous approximations:
\beq
\Gamma_{D^{*0}\rar D^0 \gamma}\simeq (8.0\pm 2.7)~\mbox{keV}
{}~~~~~~~~~~~~~~~
{\Gamma_{D^{*-}\rar D^- \gamma}}\simeq (.01\pm 0.08)~\mbox
{keV}~.
\eeq
If one instead works with the semi-local duality like-sum rule
using the complete expressions of the perturbative
contributions,
one finds that the heavy quark contribution to the coupling is
reduced
by 30\% and leads to:
\beq
\Gamma_{D^{*0}\rar D^0 \gamma}\simeq (6.7\pm 2.4)~\mbox{keV}
{}~~~~~~~~~~~~~~~
{\Gamma_{D^{*-}\rar D^- \gamma}}\simeq (.04\pm 0.08)~\mbox
{keV}~.
\eeq
Therefore, the {\it most conservative} sum rule estimate is:
\beq
\Gamma_{D^{*0}\rar D^0 \gamma}\simeq (7.3\pm 2.7)~\mbox{keV}
{}~~~~~~~~~~~~~~~
{\Gamma_{D^{*-}\rar D^- \gamma}}\simeq (.03\pm 0.08)~\mbox
{keV}~.
\eeq
which, despite the large error,
 shows that the heavy quark contribution acts in the right
direction for
explaining the large charge dependence of the observed decay
rates
\cite{PDG}. One can combine these results with the ones in
Eq. (32), for
an attempt to deduce the ratios of rates:
\beq
{\Gamma_{D^{*0}\rar D^0 \pi^0}}/
{\Gamma_{D^{*0}\rar D^0 \gamma}}~~~~~~~~~~~~
{\Gamma_{D^{*-}\rar D^0 \pi-}}/
{\Gamma_{D^{*-}\rar D^- \gamma}},
\eeq
but, due to the large errors, the comparison of the
predictions with the data is not very
conclusive. Alternatively, we can combine the predictions in
Eq. (46)
with the observed branching ratios given in \cite{PDG}. Then, we
predict:
\beq
\Gamma_{D^{*0}\rar D^0 \pi^0}\simeq (14\pm 5)~\mbox{keV}~~~~~~~
{}~~~~~~~
\Gamma_{D^{*-}\rar D^0 \pi^-}\leq 18 ~\mbox{keV}.
\eeq
and:
\beq
\Gamma_{D^{*0}\rar all}\simeq (20\pm 7) ~\mbox{keV}
{}~~~~~~~~~~~~~~
\Gamma_{D^{*-}\rar all}\leq 27  ~\mbox{keV}
{}~.
\eeq
These results are respectively in fair agreement within the
errors
 with the direct calculation in
Eq. (32) and with the prediction in Eq. (34), though the ones
for the
$D^{*-}$ have a large error due to the inaccuracy of the
measured
and predicted $D^{*-}\rar D^-\gamma$ branching ratio. The
agreements between the different results given in this paper
is an indication for the self-consistency of the
whole
approach.
\section*{Conclusion}
We have systematically studied the couplings $P^*P\pi$ and
$P^*P\gamma$
($P\equiv B,D$) using a 1$/M_b$-expansion in full QCD with the
help
of moments sum rules. It is important to notice that, like other
heavy-to-heavy transitions, the couplings are dominated by the
$hard$
perturbative diagram. This is not the case of the $B\rar\pi
(\rho)~l\nu$
and $B\rar K^*\gamma$ heavy-to-light transitions which are
governed
by the $soft$ light quark vacuum condensate \cite{SN1}. Technically, this
difference is mainly due to the uses of different
dispersion variables for the heavy-to-heavy and heavy-to-light
transition processes. We find that, for the $P^*P\pi$ couplings,
the $1/M_b$-corrections due mainly to the perturbative graph
are large but they tend to cancel for the quantity
$f_Pg_{P^*P\pi}$
and implies to a good approximation the relation in Eq. (31).
For the $P^*P\gamma$-coupling, the $1/M_b$-correction is due
mainly to
the perturbative and light quark condensate contributions from
the heavy
quark
component of the electromagnetic current, which
goes in the good direction for explaining the large charge
dependence
of the
ratio of the $ D^{*0}\rar D^0 \gamma$ over the $D^{*-}\rar
D^- \gamma$
observed widths.
\nin
For the $B^*$-mesons, our predictions are given in Eqs. (30),
(42) and (43), where for experimental interests in the next
$B$-factory
machine:
\bea
\Gamma_{B^{*-}\rar B^- \gamma}&\simeq& (.10\pm .03)~\mbox
{keV}~,\nnb\\
{\Gamma_{B^{*-}\rar B^- \gamma}}/
{\Gamma_{B^{*0}\rar B^0 \gamma}}&\simeq& 2.5~,
\eea
where the latter deviates strongly from the na\"{\i}ve static
limit ($M_b \rar \infty$) expectation $(e_u/e_d)^2=4$.
 By combining the previous different
results of the $D^*$-meson, our {\it averaged predictions}
for the different
exclusive  widths  are:
\beq
\Gamma_{D^{*-}\rar D^0 \pi^-}\simeq 1.54\Gamma_{D^{*0}\rar
D^0 \pi^0}
\simeq (8\pm 5) ~\mbox{keV}.
\eeq
and:
\bea
\Gamma_{D^{*0}\rar D^0 \gamma}&\simeq& (3.7\pm 1.2)~\mbox{keV}
{}~~~~~~~~~~~~~~~
{\Gamma_{D^{*-}\rar D^- \gamma}}\simeq (.09^{+.40}_{- 0.07})~
\mbox{keV}~.
\nnb\\
\Gamma_{D^{*0}\rar all}&\simeq& (11\pm 4) ~\mbox{keV}
{}~~~~~~~~~~~~~~
{}~~~~~~\Gamma_{D^{*-}\rar all} \simeq (12\pm 7)~\mbox{keV}~.
\eea
The branching ratios agree within the errors with the present
data though the total widths are well below the experimental
upper
limits
$\Gamma_{D^{*-}\rar all}\leq 131~\mbox{keV}$ and
$\Gamma_{D^{*0}\rar all}\leq 2~\mbox{MeV}$ \cite{PDG}.
 We urge experimentalists
to improve the measurements of these total widths in the near
future, as these measurements are necessary for clarifying the
present disagreements between different theoretical predictions.
\section*{Acknowledgements}
One of us (H.G.D) would like to thank the CNRS
for a financial support and for
the hospitality at the Laboratoire de Physique
Math\'ematique de Montpellier

\vfill\eject

\end{document}